\documentclass[aps,pra,twocolumn,superscriptaddress,10pt,noshowpacs]{revtex4}
\usepackage[english]{babel}
\usepackage[T1]{fontenc}
\usepackage[utf8]{inputenc}
\usepackage{graphicx,epstopdf}
\usepackage{amssymb}
\usepackage{amsmath}

\usepackage{amsfonts}
\usepackage{bbm}
\usepackage{color}
\usepackage{latexsym}
\usepackage{caption}
\usepackage{subcaption}
\usepackage{times,txfonts}
\usepackage{color,soul}
\usepackage{listings}
\usepackage{bbold}
\usepackage{xcolor}
\usepackage{comment}
\usepackage{natbib}
\newcommand{\beq}{\begin{equation}}
\newcommand{\eeq}{\end{equation}}
\newcommand{\bea}{\begin{eqnarray}}
\newcommand{\eea}{\end{eqnarray}}
\begin{document}

\title{Polymer Black Hole Surrounded by Quintessence}

\author{M. C. Araújo}
\email{michelangelo.araujo@ufca.edu.br}
\affiliation{Universidade Federal do Cariri, Centro de Ci\^encias e Tecnologia, 63048-080, Juazeiro do Norte, CE, Brazil}

\author{J. G. Lima}
\email{junior.lima@uaf.ufcg.edu.br}
\affiliation{Departamento de F\'{\i}sica, Universidade Federal de Campina Grande,\\
Caixa Postal 10071, 58429-900, Campina Grande, Para\'{\i}ba, Brazil}

\author{C. R. Muniz}
\email{job.furtado@ufca.edu.br}
\affiliation{Universidade Estarual do Ceará, FECLI, 63.502-253, Iguatu, CE, Brazil}

\author{J. Furtado}
\email{job.furtado@ufca.edu.br}
\affiliation{Universidade Federal do Cariri, Centro de Ci\^encias e Tecnologia, 63048-080, Juazeiro do Norte, CE, Brazil}

\begin{abstract}
In this paper, we study the polymer black hole solution surrounded by a quintessence field. The influence of quintessence on the polymer black hole is investigated through its thermodynamic properties, such as the Hawking temperature, entropy, and specific heat, which allow us to address the question of thermodynamic stability. We then calculate bounds on the electromagnetic greybody factors and photon emission rates of the black hole, highlighting the interplay between quintessence and quantum gravity effects in determining these phenomena. We also examine the effects of quintessence and quantum gravity on the geodesics and shadows of massless particles around the black hole. Our results are further compared with observational data of the Sagittarius A black hole from the Event Horizon Telescope (EHT) collaboration.
\end{abstract}

\maketitle

%\linenumbers

\section{Introduction}

As is well known, the solutions to Einstein’s field equations in General Relativity (GR) are plagued by the presence of undesirable singularities \cite{penrose1965gravitational,hawking1966properties}. These singularities mark the breakdown of the theory, indicating the limits of its applicability when quantum gravitational effects become significant—typically when the curvature reaches the Planck scale \cite{bodendorfer2019effective,Bronnikov:2024izh,Chen:2024sbc}. The actual existence of such singularities in nature is generally questioned, and it is widely believed that a complete theory of quantum gravity would be able to resolve them, replacing the singular behavior with some form of regular phenomenon. In this context, numerous approaches to quantum gravity have been proposed, among which we highlight loop quantum gravity (LQG) \cite{bojowald2007singularities,bojowald2001absence,Ashtekar:2011ni,wilson2018loop,Olmedo:2017lvt}, string theory \cite{natsuume2001singularity,Balasubramanian:2002ry,Cornalba:2002fi,cornalba2002resolution,berkooz2003comments}, the AdS/CFT correspondence \cite{maldacena2003eternal,fidkowski2004black,Craps:2007ch,hanada2014holographic,Engelhardt:2016kqb,Engelhardt:2015gla}, and frameworks involving noncommutative geometry \cite{madore2000resolution,maceda2004can,maceda2004resolution,Gorji:2014pka}.

Although a complete quantum theory of gravity has not yet been achieved, the development of numerous effective models has served as an important testing ground for addressing such classical issues. Within the context of loop quantum gravity (LQG), for instance, the successful Friedmann–Lemaître–Robertson–Walker (FLRW) cosmological models have been proposed \cite{Ashtekar:2011ni,Ashtekar:2006wn,Ashtekar:2009vc,ashtekar2015loop,Oriti:2016ueo,agullo2014loop,ashtekar2017loop,Diener:2017lde}. These models describe a homogeneous and isotropic universe on large scales and naturally resolve the singularity problem by introducing an effective spacetime in which the traditional Big Bang is replaced by a Big Bounce scenario. It is important to note that these effective models rely on the regularization of phase space through a procedure known as polymerization \cite{Corichi:2007tf}.

Recently, a quantum extension of the Schwarzschild black hole based on the polymer quantization approach of LQG was proposed \cite{Bodendorfer:2019nvy,Bodendorfer:2019cyv}. In this framework, the effective quantum spacetime gives rise to what is referred to as a polymer black hole, in which the classical singularity is replaced by a quantum bounce connecting a black hole region to a white hole region. The rotating counterpart of the polymeric Schwarzschild black hole has also been investigated in Ref. \cite{Brahma:2020eos}. It is worth mentioning that black holes inspired by LQG have been tested against observations from the Event Horizon Telescope (EHT) collaboration \cite{Islam:2022wck,Afrin:2022ztr}, with their parameters constrained by experimental data from the supermassive black holes M87* and Sgr A* \cite{Vagnozzi:2022moj,Tsukamoto:2023rzd,KumarWalia:2022ddq,Yan:2022fkr}. Additional phenomenological studies involving black hole observations and gravitational waves can be found in Refs. \cite{Brahma:2020eos,Modesto:2009ve,Alesci:2011wn,Chen:2011zzi,Dasgupta:2012nk,Barrau:2014yka,Sahu:2015dea,Cruz:2015bcj,pugliese2020constraining,Pawlowski:2014nfa,Barrau:2011md,Yang:2023gas,Fu:2023drp,Addazi:2021xuf,Garcia-Chung:2020zyq,Garcia-Chung:2022pdy,Garcia-Chung:2023oul}. 

In this paper, we analyze the polymer black hole surrounded by quintessence. Associated with negative-pressure states that drive the accelerated expansion of the universe \cite{SupernovaCosmologyProject:1998vns,SupernovaSearchTeam:1998fmf,SupernovaSearchTeam:1998cav}, quintessence emerges as an alternative proposal to dark energy, in contrast to the cosmological constant model. The latter implies an energy scale vastly larger than the one inferred from observations of dark energy \cite{Weinberg:1988cp}. In general, quintessence is described by a canonical scalar field minimally coupled to gravity, and it represents the simplest scenario capable of avoiding theoretical issues such as ghost and Laplacian instabilities (see Ref. \cite{tsujikawa2013quintessence} and the references therein for a detailed review on the topic). Here, the influence of quintessence on the polymer black hole will first be investigated in the context of black hole thermodynamics. Specifically, we focus on computing the Hawking temperature, entropy, and associated heat capacity in order to address questions related to thermodynamic stability. Subsequently, we will examine the effect of quintessence on geodesics and shadows of massless particles around the black hole. Related studies along similar lines can be found in Refs. \cite{Zeng:2020vsj,Belhaj:2020rdb,Mustafa:2022xod,Chen:2022dap,Toshmatov:2015npp}.

This paper is organized as follows: In the next section we present a brief review about the Polymer black hole, discussing the main aspects about the event horizon in such model. In section III we address the inclusion of quintessence on the Polymer black hole. In section IV we investigate the thermodynamic aspects, with focus on the Hawking temperature, heat capacity and entropy. In section V we study bounds in the electromagnetic greybody factor and particle emission rate. In section VI we investigate geodesics for light-like particles and shadows and in section VII we present our conclusions.

\section{Polymer Black Hole} \label{2}
In this section, we briefly discuss some of the features of a polymer black hole, which arises within the framework of Loop Quantum Gravity (LQG) under the condition \cite{Bodendorfer:2019nvy,Bodendorfer:2019cyv,Brahma:2020eos,Tu:2023xab}:
\begin{equation} 
ds^2 = -8A_{\lambda}\, M_{B}^{2}\, \mathcal{A}(r)\, dt^2 + \frac{dr^2}{8A_{\lambda}\, M_{B}^{2}\, \mathcal{A}(r)} + \mathcal{B}(r)\, d\Omega^2.\label{eq1} \end{equation}
This solution, which represents a quantum extension of the Schwarzschild metric in LQG, is characterized by the coefficients
\begin{eqnarray}
\mathcal{A}(r) &=& \frac{1}{\mathcal{B}(r)}\left(1 + \frac{r^2}{8A_{\lambda}\, M_{B}^{2}}\right) \left(1 - \frac{2M_B}{\sqrt{8A_{\lambda}\, M_{B}^{2} + r^2}}\right), \label{eq2} 
\end{eqnarray}
and
\begin{eqnarray}
    \mathcal{B}(r) &=& \frac{512A_{\lambda}^{3}\, M_{B}^{4}\, M_{W}^{2} + \left(r + \sqrt{8A_{\lambda}\, M_{B}^{2} + r^2}\right)^6}{8\sqrt{8A_{\lambda}\, M_{B}^{2} + r^2}\left(\sqrt{8A_{\lambda}\, M_{B}^{2} + r^2} + r\right)^3}, \label{eq3}
\end{eqnarray}
where $M_B$ and $M_W$ are Dirac observables representing conserved quantities in the model and
\begin{eqnarray}
    A_{\lambda} \equiv \frac{1}{2} \left( \frac{\lambda}{M_B\, M_W} \right)^{2/3}.
    \label{alamblambmbmw}
\end{eqnarray} 
Here, $\lambda$ is a quantum parameter associated with holonomy modifications in LQG \cite{Bodendorfer:2019nvy,Bodendorfer:2019cyv}.

As a remarkable feature, it is worth highlighting that the singularity typically associated with a polymer black hole can be replaced by a spacelike transition surface that smoothly connects an asymptotically Schwarzschild black hole to a white hole, with $M_B$ and $M_W$ denoting their respective masses. This transition occurs when the coefficient $\mathcal{B}(r)$ reaches its minimum at $r=0$ \cite{Bodendorfer:2019nvy,Bodendorfer:2019cyv,Brahma:2020eos}, thus recalling the quantum bounce scenario of the loop quantum cosmology (LQC).

\begin{figure*}
    \centering
\includegraphics[scale=0.65]{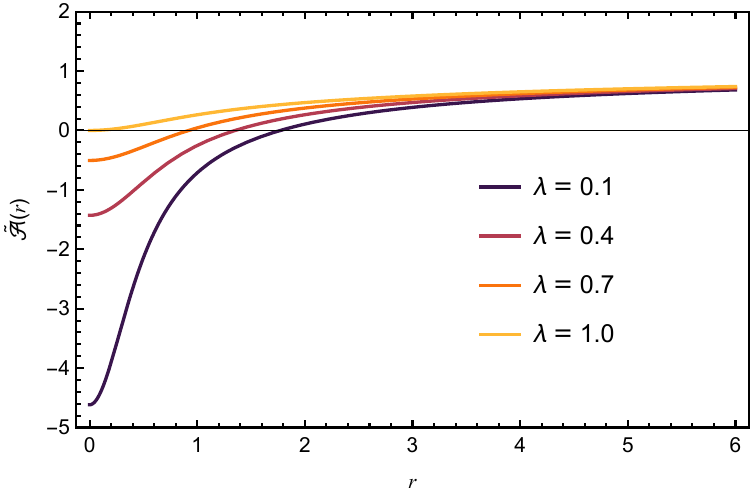}
\caption{Lapse function plotted against the radial coordinate $r$ for several values of $\lambda$. We have assumed $M=1$ for this plot.} \label{polymer111}
\end{figure*}

Now, considering a symmetric bounce structure where $M_B = M_W = M$, we can set 
\begin{equation}\label{eq6}
ds^2 = -\tilde{\mathcal{A}}(r)\, dt^2 + \frac{dr^2}{\tilde{\mathcal{A}}(r)} + \mathcal{B}(r)\left(d\theta^2 + \sin^2\theta \,  d\phi^2\right),  
\end{equation}
with $\tilde{\mathcal{A}}(r) = 8A_{\lambda}\, M^2\, \mathcal{A}(r)$ playing the role of the metric lapse function. Note that under this assumption, $\tilde{\mathcal{A}}(r)$ and $\mathcal{B}(r)$ take on the simplified forms 
\begin{eqnarray}\label{eq4}
\tilde{\mathcal{A}}(r) &=& \frac{4 (\lambda  M)^{2/3}-2 M \sqrt{4 (\lambda 
   M)^{2/3}+r^2}+r^2}{(\lambda  M)^{2/3}+r^2},  
\end{eqnarray}
and
\begin{eqnarray} \label{eq5}
\mathcal{B}(r) &=& r^2+M^2\, \left( \frac{\lambda}{M^2} \right)^{2/3},
\end{eqnarray}
respectively (we have explicitly written these expressions in terms of the parameter $\lambda$). In Fig. \ref{polymer111}, we have plotted the lapse function as a function of the radial coordinate $r$ for several values of $\lambda$, from which we can conclude that the metric defined by the line element in Eq. \eqref{eq6} leads to a single event horizon, since the cosmological horizon is absent in the model as $\tilde{\mathcal{A}}(r\rightarrow\infty) = 1$. The expression for the event horizon can be written as
\begin{eqnarray}
    r_h = 2 M \sqrt{1-\left(\frac{\lambda }{M^2}\right)^{2/3}},
    \label{eq7}
\end{eqnarray}
and the necessary condition for its existence is $0\leq \lambda \leq M^2$. Note that when $\lambda = 0 $, the Schwarzschild radius is recovered, as expected from the structure of the interval in Eq. \eqref{eq6}. Fig. \ref{polymer111} also indicates that the polymer black hole is a regular object, as no singularity is observed at any finite value of $r$. This indication can be further confirmed by examining the Kretschmann scalar, expressed as
\begin{widetext}
\begin{eqnarray}
    \nonumber K &=& \frac{16M^2}{\left(2 \lambda  M^2+r^2\right)^6 \left(8 \lambda  M^2+r^2\right)^3}\left(\lambda  M r^{10} \left(3 (37 \lambda +44) M-32 \sqrt{8 \lambda  M^2+r^2}\right)+256 \lambda ^6 M^{11} \left(3 (112 \lambda +59) M-170 \sqrt{8 \lambda  M^2+r^2}\right)\right.\\
    \nonumber &&\left.+192 \lambda ^5 M^9 r^2 \left(213 \sqrt{8 \lambda  M^2+r^2}-8 (39 \lambda +29) M\right)+32 \lambda ^4 M^7 r^4 \left((822 \lambda +654) M-569 \sqrt{8 \lambda  M^2+r^2}\right)\right.\\
    &&\left.+8 \lambda ^3 M^5 r^6 \left(3 (715 \lambda +564) M-1151 \sqrt{8 \lambda  M^2+r^2}\right)+12 \lambda ^2 M^3 r^8 \left((207 \lambda +178) M-86 \sqrt{8 \lambda  M^2+r^2}\right)+3 r^{12}\right),
\end{eqnarray}
\end{widetext}
which is clearly non-singular at $r=0$. In fact, in this limit, it takes the simplified form 
\begin{equation}
    K_{r\rightarrow 0} = \frac{3 (112 \lambda +59) M-340 \sqrt{2} \sqrt{\lambda } M}{8 \lambda ^3 M^5}.
\end{equation}

In particular, we will be concerned with this symmetric bounce structure, albeit in the context of a black hole surrounded by quintessence, as we shall implement in the next section.

\section{Quintessence effects}

According to Kiselev \cite{Kiselev:2002dx}, in a static and spherically symmetric configuration, the quintessence field surrounding a black hole must satisfy the condition
\begin{equation}
T^{\phi}_{\phi} = T^{\theta}_{\theta} = -\frac{1}{2}(3\omega + 1) T^{r}_{r} = -\frac{1}{2}(3\omega + 1) T^{t}_{t}.
\end{equation}
Here, $T^i_j$ ($i,j=t, r, \phi \text{ and } \theta$) denotes components of the energy-momentum tensor associated with the field, and $\omega$ is its equation-of-state parameter, which must satisfy $-1<\omega<-1/3$ in order to remain consistent with astronomical observations of an accelerating universe \cite{SupernovaCosmologyProject:1998vns,SupernovaSearchTeam:1998fmf,SupernovaSearchTeam:1998cav}. It is important to note that when $\omega=-1$, the scenario reduces to that of a cosmological constant, whereas $\omega=-1/3$ corresponds to a universe undergoing uniform expansion.

\begin{figure*}[t]
    \centering
\includegraphics[scale=0.65]{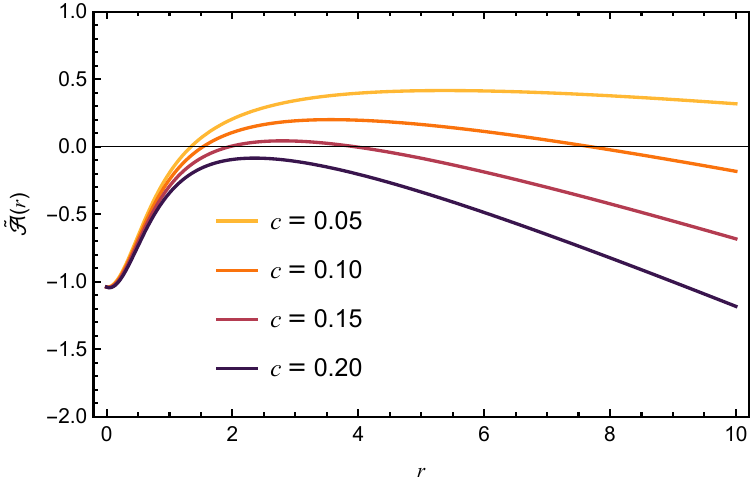}
\includegraphics[scale=0.65]{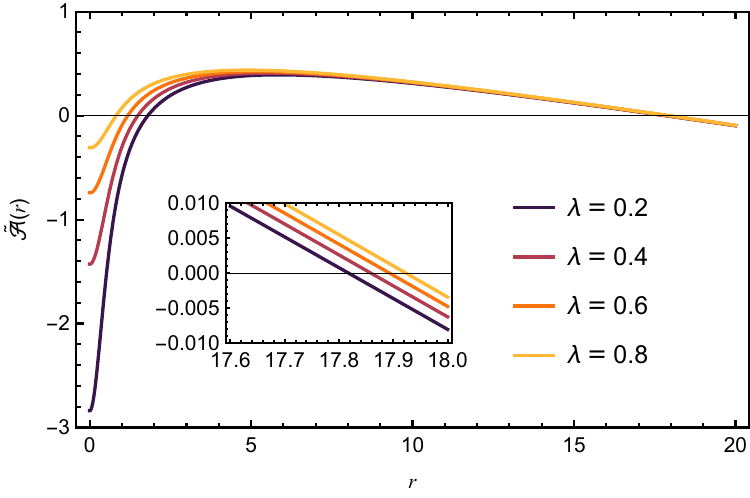}
\includegraphics[scale=0.65]{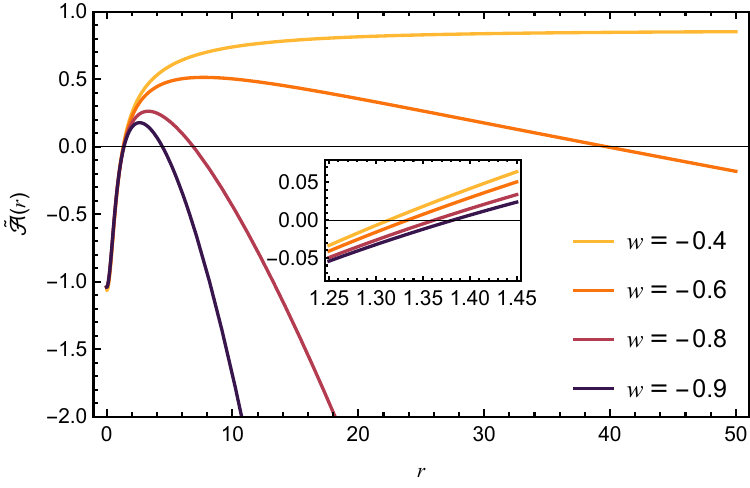}
\caption{Lapse function plotted against the radial coordinate $r$ for several values of $c$, $\lambda$ and $\omega$, respectively. In the upper left panel, we set $\lambda=0.5$ and $\omega=-2/3$. In the upper right panel, we chose $c=0.05$ and $\omega=-2/3$. In the lower panel, we considered $c=0.05$ e $\lambda=0.5$. For all plots, we assumed $M=1$.}\label{lfrclamdawm1}
\end{figure*}

In this work, we aim to investigate a combined framework involving a polymer black hole surrounded by a quintessence field. Following Kiselev's approach, the effects of quintessence can be incorporated through a redefinition of the lapse function of the metric, achieved by adding a term proportional to $r^{-3\omega-1}$. Our starting point, therefore, will be a line element of the form given in Eq. \eqref{eq6}, with a modified lapse function defined by
\begin{eqnarray}\label{lapsemetricpolymerquint}
\tilde{\mathcal{A}}(r) &=& \frac{4 (\lambda  M)^{2/3}-2 M \sqrt{4 (\lambda 
   M)^{2/3}+r^2}+r^2}{(\lambda  M)^{2/3}+r^2} - \frac{c}{r^{3\omega+1}}, \nonumber\\ 
\end{eqnarray} where $c$ is a constant parameter that effectively describes a coupling of the quintessence matter with the polymer black hole system. The behavior of Eq. \eqref{lapsemetricpolymerquint} as a function of the radial coordinate $r$ is illustrated in Fig. \ref{lfrclamdawm1}. It is evident that a direct consequence of the quintessence effect is the emergence of a new event horizon (the cosmological horizon) which was absent in the case of the pure polymer black hole, as previously discussed in the preceding section. From the upper left panel of the same figure, we can observe that increasing the value of the parameter $c$ shifts the event horizon further to the right, while simultaneously shifting the cosmological horizon further to the left. It is worth noting that the effect on the cosmological horizon is more pronounced than on the event horizon. Moreover, the condition for black hole formation (that is, the existence of a horizon) may no longer be satisfied if the value of $c$ exceeds a certain threshold, depending on the specific choice of parameters. An opposite behavior is observed in the case of the upper right panel, where the event horizon appears to shift to the left as the cosmological horizon shifts to the right for increasing values of $\lambda$. Furthermore, the impact on the former seems to be more pronounced than on the latter. It is also worth noting that the curve intersects the vertical axis at increasingly higher values as $\lambda$ grows. This suggests that, for sufficiently large values of $\lambda$, the event horizon radius ceases to exist - more precisely, it is effectively replaced by a horizon of purely cosmological nature. However, this scenario never actually occurs, since condition $0\leq \lambda \leq M^2$ must be satisfied. In the lower panel of the figure, we observe a behavior similar to that shown in the upper left panel, where the event horizon appears to shift to the right while the cosmological horizon shifts to the left. It is also worth noting that the latter seems to be significantly affected by the values of $\omega$, becoming increasingly closer to the former as $\omega$ decreases. The figure further suggests that the condition for the existence of a black hole (i.e., the existence of a horizon) might eventually be violated, since the maximum of the plotted function decreases in magnitude as $\omega$ is reduced. However, this does not actually occur due to the lower bound imposed on $\omega$, at least for the fixed parameter values considered in the figure. It is also worth noting that the presence of quintessence appears to render the solution non-regular in the limit as $r\rightarrow\infty$, as well as at the origin, since $-1<\omega -1/3$. This behavior is confirmed by the Kretschmann scalar, which in this limit ($r\rightarrow 0$) takes the form 
\begin{widetext}
\begin{eqnarray}
 K_{r \rightarrow0} &=& \frac{1}{4\lambda^3}\bigg( 512 \bigg(\frac{\lambda}{M^2}\bigg)^{2/3}(M \lambda)^{1/3}(M-2(M \lambda)^{1/3})+\lambda\bigg(49+528\bigg(\frac{\lambda}{M^2}\bigg)^{2/3}-\frac{84\lambda}{(M\lambda)^{2/3}}+\frac{36(M\lambda)^{2/3}}{M^2}\bigg)\nonumber\\
 &+& 4c^2\lambda r^{-6(1+\omega)}\bigg((2+9\omega (1+\omega ))^2\lambda^2+4(1+9\omega^2)\bigg(\frac{\lambda}{M^2}\bigg)^{2/3}r^4\bigg)+\frac{4 c \lambda r^{-3(1+\omega)}}{M^2}\bigg(-7M^2(2+9\omega(1+\omega))\lambda\nonumber\\
 &+& 6(2+9\omega (1+\omega))(M\lambda)^{4/3}+\frac{r^2}{\bigg(\frac{\lambda}{M^2}\bigg)^{1/3}}\bigg(32(-1+3\omega)\lambda+32(M\lambda)^{2/3}-96\omega(M\lambda)^{2/3}+21(1+3\omega)r^2+8\bigg(\frac{\lambda}{M^2}\bigg)^{1/3}r^2\bigg)\bigg).\nonumber\\
\end{eqnarray}
\end{widetext}
As we can see, the solution becomes regular when there is no quintessence, {\it i.e.}, for $c=0$. The full expression of the Kretschmann scalar is excessively long and, for the sake of clarity and conciseness, will not be included here.

%------------------------------

\section{Thermodynamics} \label{3}

In this section, we investigate the thermodynamic properties of the polymer black hole surrounded by quintessence, with particular emphasis on the Hawking temperature $T_H$, entropy $S$, and heat capacity $C$. This Hawking temperature, associated with the surface gravity at the event horizon, characterizes the black hole's thermal radiation as predicted by semiclassical gravity. By analyzing how \( T_H \) varies with the horizon radius under the influence of the polymerization parameter \( \lambda \) and the quintessence coupling \( c \), we gain insight into the combined impact of quantum gravity effects and dark energy components on the evaporation and thermodynamic behavior of the black hole.
\begin{figure*}
    \centering
    \includegraphics[scale=0.65]{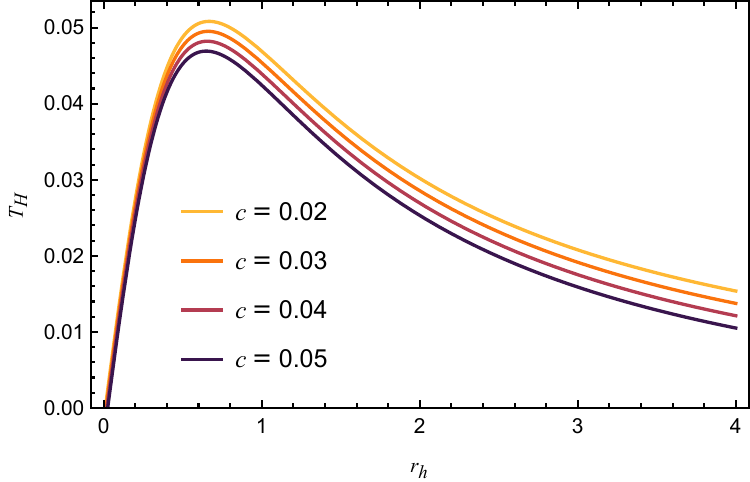}
    \includegraphics[scale=0.65]{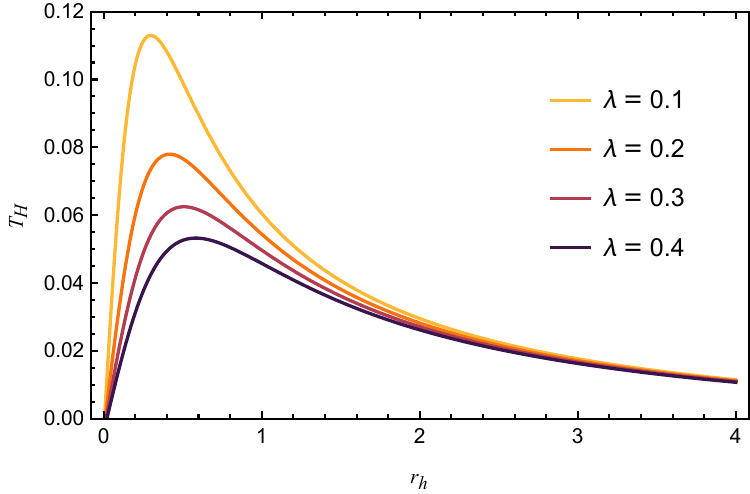}
    \caption{Plot of the Hawking temperature as a function of the event horizon radius for several values of $c$ and $\lambda$, respectively. In the left panel, we set $\lambda=0.5$. In the right panel, we chose $c=0.05$. In both cases, we have used $\omega=-2/3$.}
    \label{fig.celio2}
\end{figure*}
A detailed analysis of Fig. \ref{fig.celio2} reveals a complex interaction between the effects of quintessence and the quantum gravity corrections represented by the polymerization parameter $\lambda$. The figure suggests more than just individual parameter influences; it indicates a competition between these physical phenomena in shaping the black hole’s thermodynamic evolution. The most immediate and convergent effect observed in both panels is that an increase in either the quintessence parameter $c$ or the polymerization parameter $\lambda$ results in a decrease of the maximum Hawking temperature. The mechanism for quintessence involves a negative pressure that counteracts gravity, effectively lowering the surface gravity at the event horizon  and thus resulting in a cooler black hole. 

On the other hand, the polymer parameter $\lambda$ quantifies quantum corrections from LQG that introduce a quantum repulsion, which ``smears out'' the spacetime curvature. This also leads to a reduced surface gravity and, consequently, a lower Hawking temperature. Therefore, both phenomena (one classical and cosmological, and the other quantum-gravitational) contribute to increasing the black hole's thermodynamic stability by slowing its evaporation rate. However, a more subtle and fundamental divergence is revealed in how $c$ and $\lambda$ influence the characteristic scale, i.e., the event horizon radius $r_h$ at which this peak temperature is achieved. An increase in $c$ shifts the temperature peak to smaller horizon radii, suggesting that in a quintessence-dominated scenario, the black hole reaches its hottest emission phase when it is comparatively smaller. In contrast, an increase in $\lambda$ shifts the peak temperature to larger horizon radii. This indicates that the quantum effects become thermally dominant at a larger scale, pushing the peak emission outwards. Such a competition of scales implies that the final thermodynamics and lifecycle of such a black hole would depend on the relative strength between the influence of surrounding dark energy and the intrinsic quantum-gravitational modifications to spacetime.

\begin{figure*}
    \centering
    \includegraphics[scale=0.65]{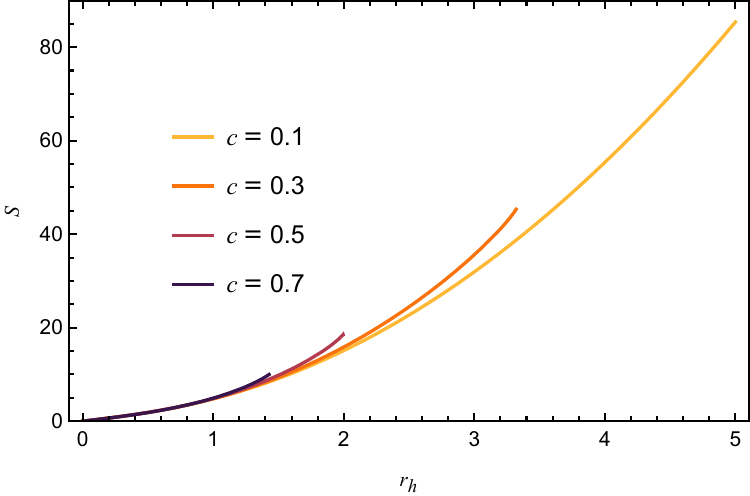}
    \includegraphics[scale=0.65]{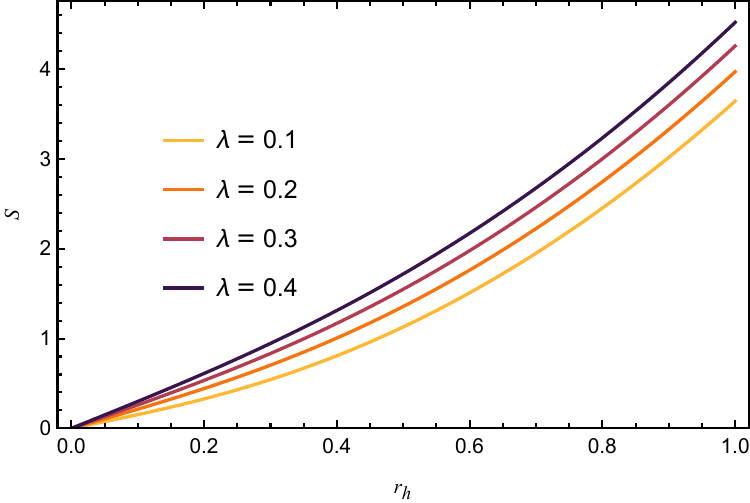}
    \caption{Plot of the entropy as a function of the event horizon radius for several values of $c$ and $\lambda$, respectively. In the left panel, we set $\lambda=0.5$. In the right panel, we chose $c=0.05$. In both cases, we have used $\omega=-2/3$.}
    \label{entropyclrh}
\end{figure*}

Fig. \ref{entropyclrh} presents the entropy as a function of the event horizon radius $r_h$,  following the prescription 
\begin{eqnarray}
    S=\int \frac{dM}{T_H}.
\end{eqnarray}
In general, one can observe that the entropy increases as either the quintessence coupling parameter $c$ or the polymerization parameter $\lambda$ increases. Furthermore, it is evident from the left panel of the figure that entropy is not well-defined for all values of $r_h$. In fact, the maximum value of the horizon radius for which the entropy remains well-behaved decreases as $c$ increases. A similar behavior is also observed in the configuration shown in the right panel, although it is not immediately evident due to the way the results were presented for clarity. There, the limiting values of $r_h$ for which the entropy is well-defined are very close to one another, indicating that they are only slightly affected by variations in $\lambda$.

\begin{figure*}
    \centering
    \includegraphics[scale=0.65]{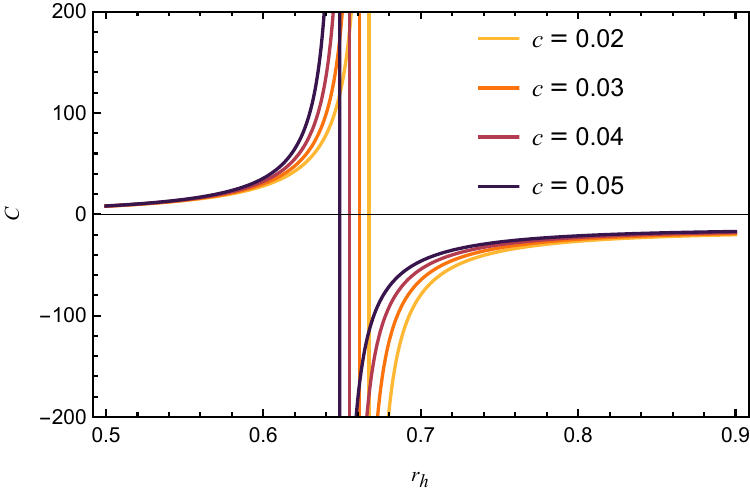}
    \includegraphics[scale=0.65]{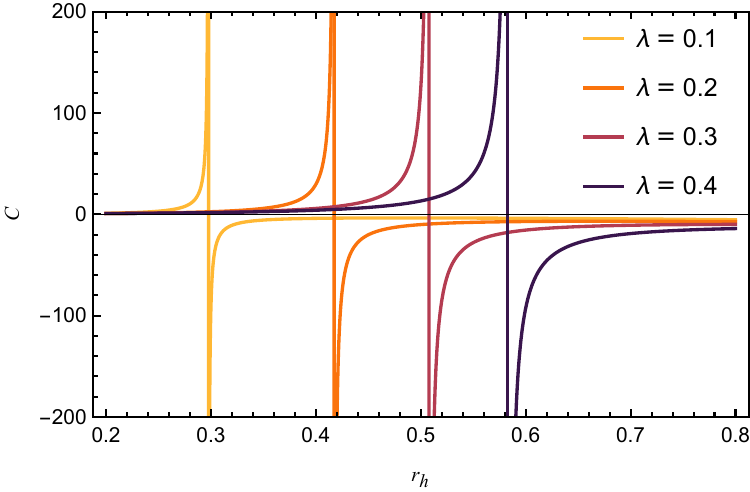}
    \caption{Plot of the heat capacity as a function of the event horizon radius for several values of $c$ and $\lambda$, respectively. In the left panel, we set $\lambda=0.5$. In the right panel, we chose $c=0.05$. In both cases, we have used $\omega=-2/3$.}
    \label{crhcandl}
\end{figure*}

With the entropy at hand, we can directly compute the heat capacity $C$, where
\begin{eqnarray}
    C = T\, \frac{\partial S}{\partial T}.
\end{eqnarray}
The results are shown in Fig. \ref{crhcandl}. The heat capacity is a fundamental thermodynamic quantity that provides insights into the local stability of the system, as it reflects how the black hole responds to small temperature perturbations. As shown in both the left and right panels of Fig. 5, one can identify stable states (characterized by positive values of $C$) and unstable states (negative values of $C$), associated with second-order phase transitions marked by discontinuities in the maximum value of the Hawking temperature. Furthermore, for both the stable and unstable configurations in the left panel, the heat capacity appears to increase as the parameter $c$ increases. An opposite behavior is observed in the right panel, where the entropy is higher for smaller values of $\lambda$ in both configurations of the system.

\section{Bounds in the electromagnetic greybody factor and particle emission rate}\label{gray}

In this section, we calculate the greybody factor lower bounds associated with the transmission of outgoing photons through the effective potential barrier surrounding the black hole. These bounds can also be expressed in terms of Bogoliubov coefficients. Such results may be considered as general bounds on these coefficients. Due to their generality, they serve as powerful tools for gaining qualitative insight into one-dimensional scattering phenomena. Moreover, this approach provides a unifying framework that connects a variety of seemingly unrelated results derived in specific cases. For more details, see Refs. \cite{Visser:1998ke,Boonserm:2023oyt}.

The radial equation for the electromagnetic field can be written in terms of a Schrödinger-like equation by introducing tortoise coordinates, leading to:
\begin{equation}
    \frac{d^2 \mathcal{R}}{dr_{*}^2}+[\omega_*^{2}+V_{eff}(r(r_*))]\mathcal{R}=0,
\end{equation}
where $\omega_*$ is the frequency, identified with the particle energy. The tortoise coordinate is defined from $dr_*/dr=\tilde{\mathcal{A}}(r)^{-1}$ and the effective potential is written as (see Ref. \cite{Arbey:2021yke}) 
\begin{equation}\label{effpotgrey}
V_{eff}(r)=\ell(\ell+1) \frac{\tilde{\mathcal{A}}(r)}{\mathcal{B}(r)}.
\end{equation}

Here, we consider the mode $\ell=1$. The lower bound for the greybody factor \(\Gamma_{\ell}^{\omega_*}\), identified with the transmission coefficient through the potential barrier, which quantifies the deviation of a black hole’s radiation from that of an ideal black body, is given by (see Ref. \cite{Al-Badawi:2023emj})
\begin{equation}
    \Gamma_{\ell}^{\omega_*}\geq\text{sech}^2\left(\int_{r_h}^{\infty}\frac{V_{eff}}{2\omega_*}dr_{*}\right),
\end{equation}
and the particle emission rate (number of particles per time unit per frequency unit) is 
\begin{equation}
    \frac{d^2 N}{dtd\omega_*}=\frac{1}{2\pi}\frac{\Gamma_{\ell}^{\omega_*}}{[\exp{\left(\omega_*/T\right)}-1]},
\end{equation}
where $T$ is the absolute temperature \cite{Arbey:2021yke}, that in our case coincides with the Hawking one.

Considering the electromagnetic effective potential in Eq. (\ref{effpotgrey}) as well as the expressions (\ref{eq5}) and (\ref{lapsemetricpolymerquint}), we arrive at
\begin{equation}
\Gamma_{\ell}^{\omega_*}\geq\text{sech}^2\left[\frac{\ell (\ell+1)}{2\omega} \left(\frac{1}{2} \pi  \sqrt{\frac{1}{\mathcal{B}(r_h)}}-\frac{\tan ^{-1}\left(\frac{r_h}{\sqrt{\mathcal{B}(r_h)}}\right)}{\sqrt{\mathcal{B}(r_h)}}\right) \right].
\end{equation}
\begin{figure*}[t]
    \centering
    \includegraphics[scale=0.65]{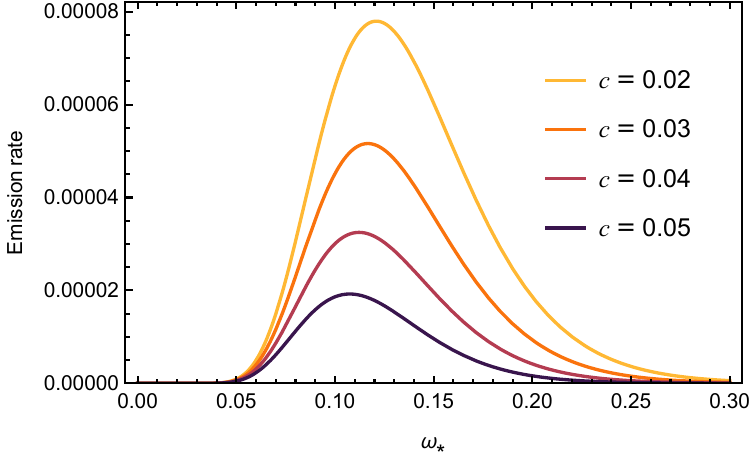}
    \includegraphics[scale=0.65]{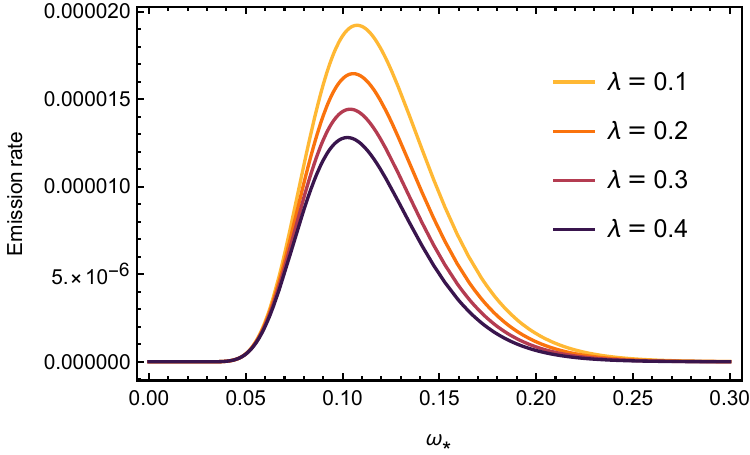}
    \includegraphics[scale=0.65]{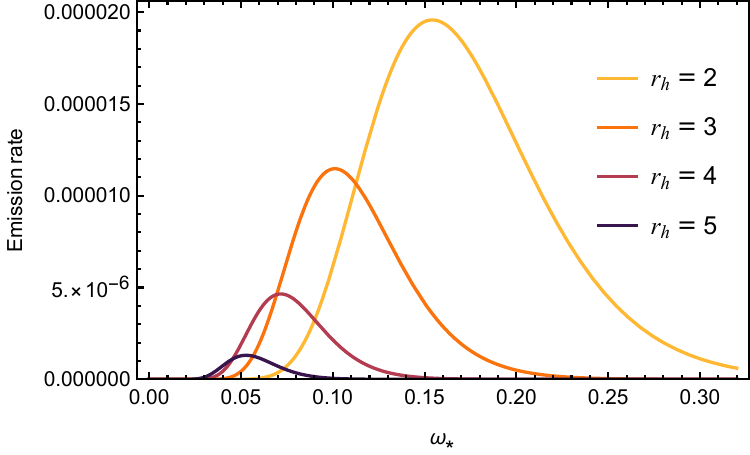}
    \caption{Photon emission rate as a function of the particle energy $\omega_*$ for selected values of $c$, $\lambda$ and $r_h$, respectively. In the upper left panel, we set $\lambda=0.1$ and $r_h=3$. In the upper right panel, we chose $c=0.05$ and $r_h=3$.  In the lower panel, we considered $c=0.05$ and $\lambda=0.5$.  In all cases, we have used $\ell=1$ and $\omega=-2/3$.}
    \label{fig.celio3}
\end{figure*}
The analysis of the lower bound for the photon emission rate, depicted in Fig. \ref{fig.celio3}, as a function of particle energy $\omega_*$, reveals the distinct thermodynamic impact of quintessence and polymer parameters. The upper panels demonstrate a clear trend: an increase in either the quintessence coupling $c$ or the polymer parameter $\lambda$ leads to a significant suppression of the emission rate across the entire energy spectrum. This result is a direct consequence of the black hole's modified thermodynamics. As established in the analysis of the Hawking temperature, both a stronger quintessence field and more pronounced polymer corrections act to cool the black hole. According to the particle emission formula, a lower temperature $T_H$ increases the magnitude of the thermal term in the denominator, $[\exp\,  (\omega_*/T_H)\, -1]$, thereby reducing the number of emitted particles per unit time. This confirms that both quintessence and quantum gravity effects make the black hole a less efficient thermal radiator.

Furthermore, the lower panel of Fig. \ref{fig.celio3} illustrates the influence of the black hole's size on the emission rate by varying the event horizon radius $r_h$. The graph shows that as this latter increases, the peak emission rate diminishes drastically, and the entire emission spectrum shifts towards lower particle energies $\omega_*$. This behavior is characteristic of thermal radiation from black bodies, where larger (and typically cooler) objects radiate less intensely and have their emission peak at lower frequencies. For the given set of parameters, a larger black hole is thermodynamically cooler, which explains both the overall reduction in luminosity and the "redshift" of its peak energy output. This underscores that the horizon size is a dominant factor in governing the black hole's radiative properties, with larger black holes being significantly dimmer and emitting lower-energy photons.

\section{Geodesics and shadows}
\label{geosha}

\begin{figure*}[t]
    \centering
    \includegraphics[scale=0.5]{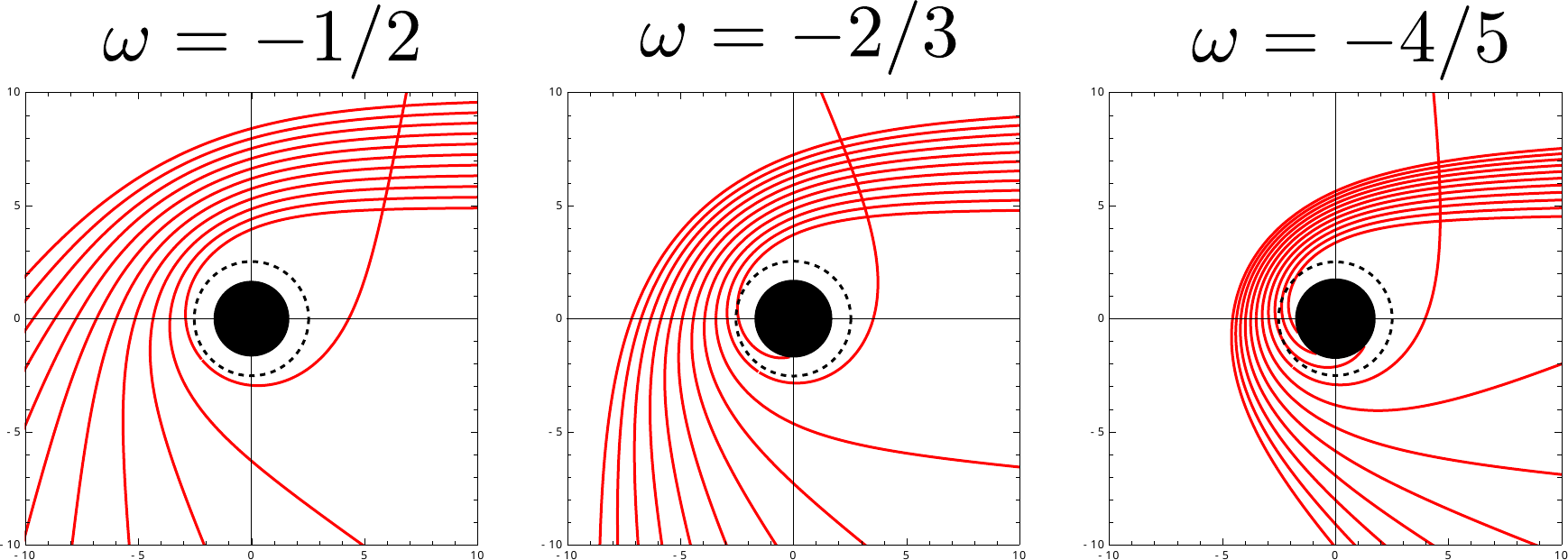}
    \caption{Geodesics for the light trajectory around a Polymer Black Hole surrounded by a Quintessence field where we aim to investigate the influence of the $\omega$ parameter in the light trajectories. For this purpose, we have set $c=0.05$, $A_\lambda=0.2$ and vary the $\omega$ parameter by choosing $\omega=-1/2$, $\omega=-2/3$ and $\omega=-4/5$.}
    \label{fig_geodesics_1}
\end{figure*}

In this section, we will analyze the geodesic motion of lightlike test particles around the Polymer black hole surrounded by quintessence. Since our metric depends only on the radial coordinate $r$ (see Eq. \eqref{eq6} and Eq. \eqref{lapsemetricpolymerquint}), two Killing vectors associated with conserved quantities can be identified. The first one, $K_\mu = (-\tilde{\mathcal{A}}(r), 0, 0, 0)$, related to time-translation symmetry, leads to a conserved energy expression:
\begin{equation}
    E = \tilde{\mathcal{A}}(r)\, \dot{t} = \text{constant},
\end{equation}
where the dot denotes differentiation with respect to the affine parameter. The second vector, $K_\mu = (0, 0, 0, \mathcal{B}(r))$, leads to the conservation of angular momentum:
\begin{equation}\label{4.12}
    L = \mathcal{B}(r)\dot{\varphi}= \text{constant}.
\end{equation}
In this analysis, we restrict ourselves to the equatorial plane, which corresponds to $\theta =\pi/2$, a simplification allowed by the spherical symmetry of the system. Under this condition, the metric given in Eq. \eqref{eq6}, can be simplified to the form
\begin{equation}\label{4.10}
    ds^2 = -\tilde{\mathcal{A}}(r)dt^2 + \frac{1}{\tilde{\mathcal{A}}(r)}dr^2 + \mathcal{B}(r)d\varphi^2.
\end{equation}

The Lagrangian function associated with the above metric takes the form
\begin{equation}
    \mathcal{L} = \dfrac{1}{2}\left(-\tilde{\mathcal{A}}(r)\dot{t}^2 + \dfrac{\dot{r}^2}{\tilde{\mathcal{A}}(r)}  + \mathcal{B}(r)\dot{\varphi}^2 \right),
\end{equation}
which can be further rearranged in terms of the derivative of the radial coordinate with respect to the affine parameter as
\begin{equation}\label{4.14}
    \dot{r}^2 = E^2 - \tilde{\mathcal{A}}(r)\left(\epsilon + \dfrac{L^2}{\mathcal{B}(r)} \right).
\end{equation}
Here, $\epsilon = -2\mathcal{L} $
represents the norm of the tangent vector along the geodesic path and takes the value zero for the lightlike geodesics of interest. Thus, we can write 
\begin{equation}\label{4.15}
    \dot{r}^2 = E^2 - V(r)
\end{equation}
where
\begin{eqnarray}
    V(r) &=& \Bigg[\frac{4 (\lambda  M)^{2/3}-2 M \sqrt{4 (\lambda 
   M)^{2/3}+r^2}+r^2}{(\lambda  M)^{2/3}+r^2}\nonumber\\
   &-& \frac{c}{r^{3\omega+1}} \Bigg]\left(\epsilon + \dfrac{L^2}{\mathcal{B}(r)} \right)
\end{eqnarray}
represents the effective potential for the Polymer black hole surrounded by quintessence. From Eqs. \eqref{4.12} and \eqref{4.15}, we finally get the following equation for null geodesics:
\begin{equation}
    \dfrac{d\varphi}{dr} = -\dfrac{b}{\mathcal{B}(r)}\left(1 - \dfrac{b^2}{\mathcal{B}(r)}\tilde{\mathcal{A}}(r)\right)^{-1/2},
\end{equation}
where $b = L/E$ defines the impact parameter.

\begin{figure*}[ht!]
    \centering
    \includegraphics[scale=0.4]{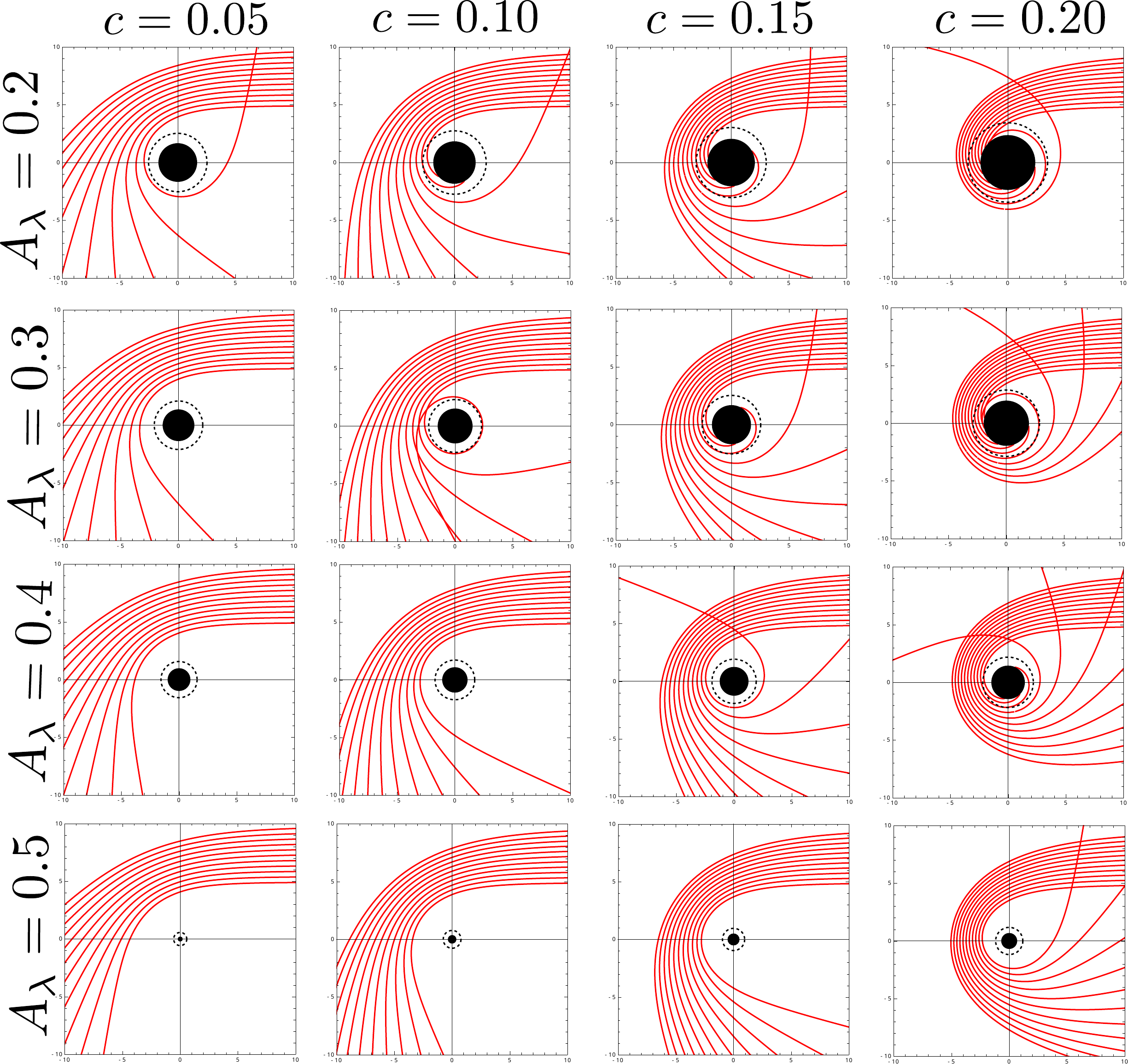}
    \caption{Geodesics for the light trajectory around a Polymer Black Hole surrounded by a Quintessence field. We have considered for this plot $\omega=-1/2$ and $M=1$. We have varied the Loop Quantum Gravity $A_\lambda$ parameter and Quintessence $c$ parameter.}
    \label{fig_geodesics_2}
\end{figure*}

In Figs. \ref{fig_geodesics_1} and \ref{fig_geodesics_2} we have depicted the ray tracing for light particles around the Polymer black hole under the influence of quintessence. For both plots we have considered $M=1$. In Fig. \ref{fig_geodesics_1} we aim to investigate the influence of the $\omega$ parameter in the light trajectories. For this purpose, we have set $c=0.05$, $A_\lambda=0.2$ (here we choose to work with $A_{\lambda}$ rather than $\lambda$, given that $M=1$. See Eq. \eqref{alamblambmbmw}) and vary the $\omega$ parameter by choosing $\omega=-1/2$, $\omega=-2/3$ and $\omega=-4/5$. As can be clearly seen from the figure, decreasing the value of $\omega$ toward $\omega = -1$ causes the black hole horizon radius to increase, and light particles tend to become more confined. It is important tho highlight that, for all plots we have considered the same impact parameter as well.

In Fig. \ref{fig_geodesics_2}, we focus on examining the geodesics of light trajectories around a Polymer Black Hole surrounded by a quintessence field, for a fixed value of $\omega$, namely $\omega = -1/2$, while varying the Loop Quantum Gravity parameter $A_\lambda$ and the quintessence parameter $c$. In this figure, we have constructed a panel in which the rows have a fixed value for $A_\lambda$ while the columns have a fixed value for $c$. We have considered four for values of $c$, namely, $c=0.05$, $c=0.10$, $c=0.15$ and $c=0.20$. Similarly, we have considered four values for $A_\lambda$, namely, $A_\lambda=0.2$, $A_\lambda=0.3$, $A_\lambda=0.4$ and $A_\lambda=0.5$. Notice that, for any value of $A_\lambda$, if we increase the value of the $c$ parameter, the black hole becomes larger, i.e., the increasing of $c$ promotes an increasing of the horizon radius, and therefore, the net effect on the light trajectory is a great tendency to confinement. On the other hand, if we consider any value of $c$, the increasing in the $A_\lambda$ parameter promotes a decreasing in the size of the black hole (smaller horizon radius) which leads to a smaller tendency to confinement of the light particles around the black hole. The values of the horizon radius $r_h$, the photon ring radius $r_{pr}$ and the shadow radius $r_{sh}$ for the $(c, A_\lambda)$ configurations according to the panel in Fig. \ref{fig_geodesics_2} can be consulted in table \ref{table_radius}.

\begin{figure*}[t]
    \centering
    \includegraphics[scale=0.7]{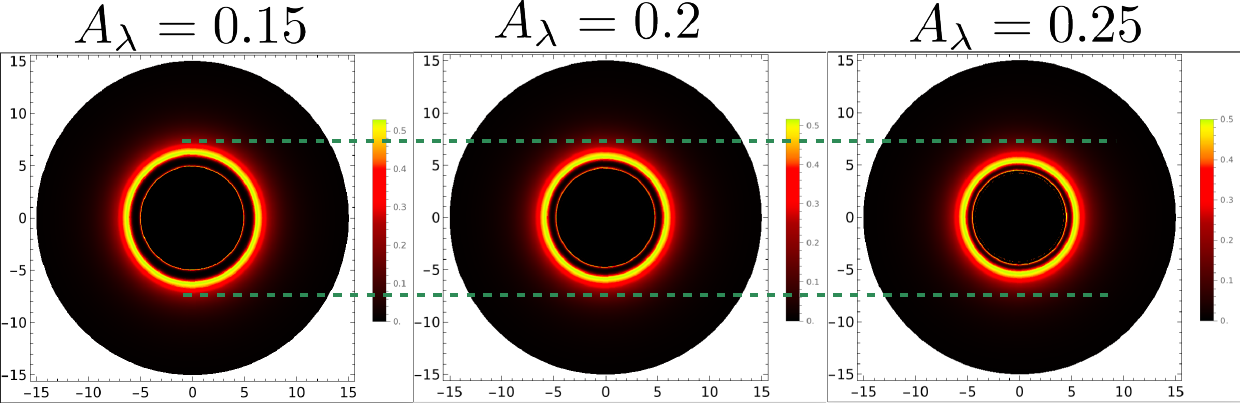}
    \caption{Shadows plots for $\omega=-2/3$ and $c=0.001$.}
    \label{shadows_1}
\end{figure*}

In Fig. \ref{shadows_1}, we have depicted the shadows for three configurations of the parameters. For the plots we have considered $\omega=-2/3$, $c=0.001$ and three values of $A_\lambda$, namely, $A_\lambda=0.15$, $A_\lambda=0.2$ and $A_\lambda=0.25$. The green dashed line was included in the figure for the sake of comparison, so that it becomes easier to see that the radius of the shadow indeed becomes smaller as we increase the value of $A_\lambda$.

We also compare our shadows results with the observational data of the Sagittarius A (Sgr A*) black hole. According to the Keck data from the EHT collaboration \cite{EventHorizonTelescopeCollaboration_2022, Vagnozzi_2023}, the image of the Sgr A* shadow is such that
\begin{equation}\label{sgrA}
    4.55\leq \frac{r_{sh}}{M}\leq 5.22.
\end{equation}

\begin{figure*}[t]
    \centering
    \includegraphics[scale=0.33]{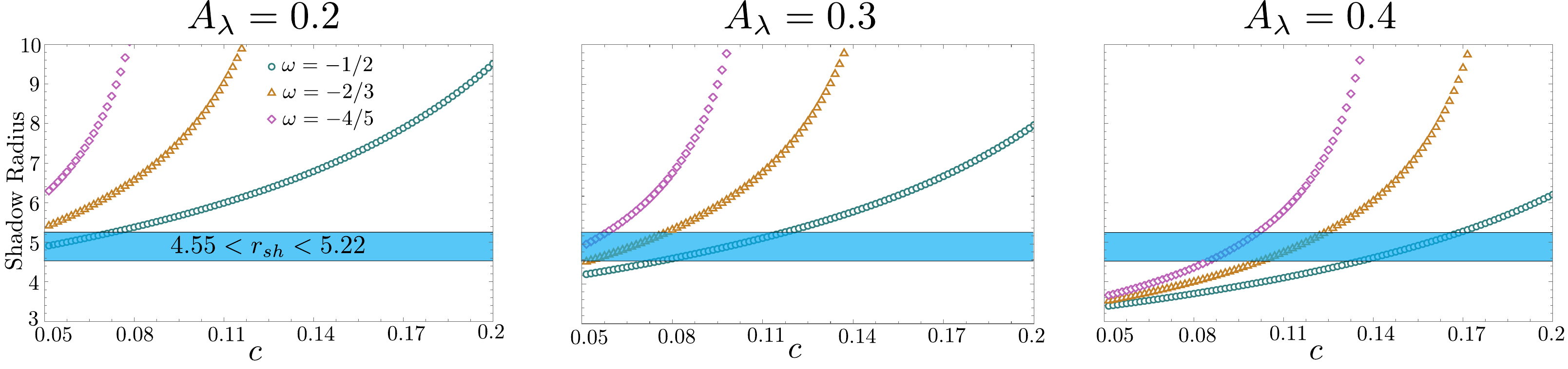}
    \caption{Comparisson of the shadow radius with the data from the Sagittarius A* black hole. Notice that there is a wide range of values of $c$, for given $(\omega, A_\lambda)$ in which the shadow radius is in the range of the SgrA* black hole.}
    \label{bounds}
\end{figure*}

In Fig. \ref{bounds}, we compared the values of the shadow radius for the Polymer black hole surrounded by quintessence with the range of the shadow of Sgr A*, expressed by Eq. \eqref{sgrA}. We considered three values of $A_\lambda$, namely, $A_\lambda=0.2$, $A_\lambda=0.3$ and $A_\lambda=0.4$, each one for each graph in Fig. \ref{bounds}. We also considered three values of $\omega$, namely $\omega=-1/2$, $\omega=-2/3$ and $\omega=-4/5$. In each graph in Fig. \ref{bounds}, we varied $c$ in a range from $0.05$ to $0.2$. It becomes clear that there is a wide range of values for $c$, in combination with $A_\lambda$ and $\omega$, such that the shadow radius for the Polymer black hole surrounded by quintessence matches the range of the image from Sgr A*. It is important to highlight that the shape and size of the shadow boundary have a very weak dependence on black hole spin and the observer’s inclination, hence our considerations of axial shadows analysis and the fact that our black hole model has no angular momentum, are in agreement with the EHT data consideration.

\begin{table}
    \centering
    \begin{tabular}{c||c|c|c|c}\hline\hline
         & $c=0.05$ & $c=0.1$ & $c=0.15$ & $c=0.2$\\\hline\hline
        $A_\lambda=0.2$ & 
        \begin{tabular}{c}
       $r_h=1.6748$ \\
       $r_{pr}=2.5271$ \\
       $r_{sh}=4.9138$ \\
        \end{tabular} & 
        \begin{tabular}{c}
       $r_h=1.8353$ \\
       $r_{pr}=2.7467$ \\
       $r_{sh}=5.8030$ \\
        \end{tabular} & 
        \begin{tabular}{c}
        $r_h=2.0528$ \\
        $r_{pr}=3.0347$ \\
        $r_{sh}=7.1669$ \\
        \end{tabular} &
        \begin{tabular}{c}
        $r_h=2.3775$ \\
        $r_{pr}=3.4409$ \\
        $r_{sh}=9.6171$ \\
        \end{tabular}\\\hline
         $A_\lambda=0.3$& 
         \begin{tabular}{c}
        $r_h=1.3734$ \\
        $r_{pr}=2.1007$ \\
        $r_{sh}=4.2420$ \\
        \end{tabular} & 
        \begin{tabular}{c}
        $r_h=1.5104$ \\
        $r_{pr}=2.2891$ \\
        $r_{sh}=4.9913$ \\
        \end{tabular} & 
        \begin{tabular}{c}
        $r_h=1.6929$ \\
        $r_{pr}=2.5344$ \\
        $r_{sh}=6.1177$ \\
        \end{tabular} &
        \begin{tabular}{c}
        $r_h=1.9585$ \\
        $r_{pr}=2.8760$ \\
        $r_{sh}=8.0618$ \\
        \end{tabular}\\\hline
        $A_\lambda=0.4$ & 
        \begin{tabular}{c}
        $r_h=0.9871$ \\
        $r_{pr}=1.5654$ \\
        $r_{sh}=3.3902$ \\
        \end{tabular} & 
        \begin{tabular}{c}
        $r_h=1.1009$ \\
        $r_{pr}=1.7200$ \\
        $r_{sh}=3.9790$ \\
        \end{tabular} & 
        \begin{tabular}{c}
        $r_h=1.2478$ \\
        $r_{pr}=1.9171$ \\
        $r_{sh}=4.8416$ \\
        \end{tabular} &
        \begin{tabular}{c}
        $r_h=1.4527$ \\
        $r_{pr}=2.1854$ \\
        $r_{sh}=6.2625$ \\
        \end{tabular}\\\hline
        $A_\lambda=0.5$ & 
        \begin{tabular}{c}
        $r_h=0.2220$ \\
        $r_{pr}=0.5696$ \\
        $r_{sh}=1.9296$ \\
        \end{tabular} & 
        \begin{tabular}{c}
        $r_h=0.3705$ \\
        $r_{pr}=0.7708$ \\
        $r_{sh}=2.3786$ \\
        \end{tabular} & 
        \begin{tabular}{c}
        $r_h=0.5200$ \\
        $r_{pr}=0.9624$ \\
        $r_{sh}=2.9590$ \\
        \end{tabular} &
        \begin{tabular}{c}
        $r_h=0.6915$ \\
        $r_{pr}=1.1778$ \\
        $r_{sh}=3.8195$ \\
        \end{tabular}\\
        \hline \hline
        \end{tabular}
    \caption{Values of the horizon radius $r_h$, the photon ring radius $r_{pr}$ and the shadow radius $r_{sh}$ for the $(c, A_\lambda)$ configurations according to the panel in Fig. \ref{fig_geodesics_2}.}
    \label{table_radius}
\end{table}

\section{Final remarks}\label{con}

In this paper, we have investigated the polymer black hole surrounded by quintessence in different contexts, such as black hole thermodynamics and the geodesics and shadows of massless particles. Our analysis in Sec. \ref{3} reveals an intriguing interplay between cosmological and quantum gravitational effects. An increased influence of dark energy, modeled by the quintessence field and parameterized by the constant $c$ (we have used the state parameter $\omega=-2/3$), leads to a decrease in the Hawking temperature, with the thermal peak occurring for smaller black holes. In contrast, a stronger presence of quantum gravity corrections, governed by the polymerization parameter $\lambda$, also results in lower temperatures, but with the thermal peak shifting toward larger black hole sizes. This contrasting behavior illustrates how each physical contribution -- cosmological in the case of quintessence and quantum in the case of polymer effects -- modulates the black hole's thermodynamic behavior in distinct ways, ultimately shaping its evaporation process and thermal evolution. In the same section, we carried out an analysis of the entropy and the associated heat capacity, where we found that the entropy is not well-defined for all values of the horizon radius $r_h$. In fact, the maximum value of $r_h$ for which the entropy remains well-defined decreases as both the cosmological parameter $c$ and the quantum parameter $\lambda$ increase. Moreover, the entropy tends to take increasingly larger values as $c$ and $\lambda$ grow. Regarding the heat capacity, our results indicate the presence of both stable and unstable states, associated with second-order phase transitions occurring at the maximum value of the Hawking temperature. For both stable and unstable configurations, the heat capacity was found to increase with increasing values of $c$. In contrast, an opposite behavior was observed with respect to the parameter $\lambda$, where smaller values of it lead to a larger heat capacity.

Following, we investigated the lower bounds for the greybody factors of the electromagnetic radiation and the associated particle emission rate. We observed that both the parameters $c$ and $\lambda$ act synergistically to suppress the particle emission spectrum, which is consistent with the previously discussed reduction in Hawking temperature. The overall decrease in the emission rate across the energy spectrum, along with the shift of the peak to lower frequencies as the event horizon radius increases, reflects the typical behavior of larger and cooler black bodies. These findings reinforce the notion that the polymeric black hole surrounded by quintessence behaves as a less efficient thermal radiator, with a significantly redshifted emission spectrum -- an effect that may have important implications for the observational signatures of such objects.

The analysis developed in Sec. \ref{geosha} investigates how the LQG parameter $\lambda$ (or $A_\lambda$, for the $M=1$ case) and the quintessence parameters $\omega$ and $c$ influence geodesics of light-like particles and the black hole shadow. We could see that, for fixed $c$ and $A_\lambda$, by decreasing the value of $\omega$, approaching $\omega=-1$, the black hole horizon radius becomes bigger and the light particles tend to become more confined. In order to investigate the role played by the $c$ and $A_\lambda$ parameter in the light-like particles geodesics, we have set $\omega=-1/2$ and varied $c$ and $A_\lambda$. We could notice that, for any value of $A_\lambda$, if we increase the value of the $c$ parameter, the black hole becomes larger, i.e., the increasing of $c$ promotes an increasing of the horizon radius, and therefore, the net effect on the light trajectory is a greater tendency to confinement. On the other hand, if we consider any value of $c$, the increasing in the $A_\lambda$ parameter promotes a decreasing in the size of the black hole (smaller horizon radius) which leads to a smaller tendency to confinement of the light particles around the black hole. Also, the shadows were computed for several configurations and we could find a wide range of configurations of $(A_\lambda, c, \omega)$ in which the shadow size matches the data of the Sagittarius A* black hole.

\section*{Acknowledgments}
\hspace{0.5cm} M. C. Ara\'{u}jo would like to thank FUNCAP, Funda\c{c}\~{a}o Cearense de Apoio ao Desenvolvimento Cient\'{i}fico e Tecnol\'{o}gico (Process No. DC3-0235-00076.01.00/24), and CNPq, Conselho Nacional de Desenvolvimento Cient\'{i}fico  e Tecnol\'{o}gico - Brasil (Process No. 304145/2025-4) for financial support. J. G. Lima would like to thank the Coordena\c{c}\~{a}o de Aperfei\c{c}oamento de Pessoal de N\'{i}vel Superior (CAPES) - Finance Code 001. C. R. Muniz would like to thank CNPq by the partial support, grant 308168/2021-6. J. Furtado would like to thank Alexandra Elbakyan and Sci-Hub, for removing all barriers in the way of science and the Funda\c{c}\~{a}o Cearense de Apoio ao Desenvolvimento Cient\'{i}fico e Tecnol\'{o}gico (FUNCAP) under the grant PRONEM PNE0112- 00085.01.00/16 and the Conselho Nacional de Desenvolvimento Científico e Tecnol\'{o}gico (CNPq) under the grant 304485/2023-3.

%%%%%%%%%%%%%%%%%%%%%%%%%%%%%%%%%%%%%%%%%%%%%%%%%%%%%%%%%%%%%%%%%%%%%%%%

%\bibliographystyle{apsrev4}
\bibliography{Ref}

%%%%%%%%%%%%%%%%%%%%%%%%%%%%%%%%%%%%%%%%%%%%%%%%%%%%%%%%%%%%%%%%%%%%%%%%
\begin{comment}

\end{comment}

\end{document}